\documentclass[amsmath,superscriptaddress,showpacs,prb,twocolumn] {revtex4}
\usepackage{bm}
\usepackage{enumerate}
\usepackage{graphicx}
\usepackage[dvips]{epsfig}
\usepackage{epsf}
\usepackage{physics}

\makeatletter
\newcommand{\Rmnum}[1]{\expandafter\@slowromancap\romannumeral #1@}
\makeatletter

\begin{document}
\title{Chiral electric separation effect in Weyl semimetals}
\author{Vladimir A. Zyuzin}
\affiliation{Department of Physics and Astronomy, Texas A$\&$M University, College Station, Texas 77843-4242, USA}
\begin{abstract}
We study the chiral electric separation effect (CESE) in Weyl semimetals (WSM). 
Within the model based on the kinetic equation we show that there is a non-zero chirality current in external electric and magnetic fields. 
We distinguish longitudinal, in-plane transverse, Hall, and anomalous Hall components of the CESE. 
It is shown that the first two components are quadratic in electric field and linear in magnetic field, while the Hall component is quadratic in both the electric and magnetic fields. 
All three are due to the chiral anomaly.
In WSMs the chirality current can be associated with the spin current, which can be experimentally observed.
\end{abstract}

\maketitle

\section{Introduction}
An understanding that condensed matter realization of the relativistic, Dirac-like, physics can occur in the narrow-gap semiconductors with strong spin-orbit coupling came from pioneering works Refs. [\onlinecite{AbrikosovBeneslavskiiJETP1971, VolkovPankratovJETP1985, NielsenNinomiyaPLB1983}], and has recently resulted in a field of Dirac and Weyl semimetals. \cite{MurakamiNJP2007,SavrasovPRB2011}

A WSM is a three-dimensional metal whose band structure necessary has a linear touching of valence and conduction bands , i.e. a degeneracy.
The degeneracy is characterized by the Berry curvature. \cite{Berry} 
Berry curvature is an effective magnetic field in momentum space the fermions propagate in. 
Formally, one can associate a chirality with the degeneracy - the way fermions spin degree (real spin or pseudospin) of freedom is aligned with its momentum: parallel ($+$ chirality) or anti-parallel ($-$ chirality). 
Distinct to WSM surface Fermi arcs \cite{SavrasovPRB2011}, chiral Landau levels \cite{NielsenNinomiyaPLB1983}, and the chiral anomaly \cite{AdlerPR1969, BellJackiwNCS1969, NielsenNinomiyaPLB1983, FukushimaKharzeevWarringaPRD2008} can be associated with the effect of the Berry curvature.

Using the kinetic equation with semiclassical equations of motions modified by the Berry curvature \cite{XiaoChangNiuRMP2010, SonYamamotoPRD2013, GorbarPRL2017}, a general magnetoconductivity tensor \cite{SeitzPR1950,ZimanBook} due to the chiral anomaly \cite{AdlerPR1969, BellJackiwNCS1969} in WSM was derived in a number of works. \cite{NielsenNinomiyaPLB1983, SonSpivakPRB2013, SpivakAndreevPRB2016, Yip, ZyuzinWSM}
This includes the positive longitudinal magnetoconductivity (LMC) \cite{SonSpivakPRB2013, SpivakAndreevPRB2016, Yip, ZyuzinWSM} and in-plane transverse magnetoconductivity (TMC) \cite{zrte5_exp1, Yip, ZyuzinWSM} (also known as planar Hall). 
Also, the chiral anomaly can result in linear magnetoconductivity given the time-reversal symmetry is broken in the crystal. \cite{CortijoPRB2016, ZyuzinWSM}

Positive LMC constitutes a unique transport signature of the WSM, and was observed in a large number of materials. \cite{cd3as2_exp1, cd3as2_exp2, cd3as2_exp3, na3bi_exp, zrte5_exp1, taas_exp1, taas_exp2}
The in-plane TMC component was recently observed in Refs. [\onlinecite{PHEexp1,PHEexp2}].    
Anisotropic linear magnetoconductivity in a presumably magnetic WSM with tilted Dirac spectrum \cite{ZyuzinWSM} was recently observed in Ref. [\onlinecite{magneticWSMexp}].

In this paper we discuss a generation of the chirality currents, first proposed in the context of high-energy physics and which can be relevant for the experiments on the WSMs. 
There is a way to make currents of fermions of opposite chiralities to either flow in opposite direction, or have different magnitudes but flow in the same direction.  
In the first case it is the chiral separation effect (CSE) \cite{SonZhitnitskyPRD2004, MetlitskiZhitnitskyPRD2005} which occurs in external magnetic field. 
In the second case it is the chiral electric separation effect (CESE) \cite{HuangLiaoCESE2013, PuWuYangPRD2014CESE, PuWuYangPRD2015CHE, GorbarCESE2016} (and see Ref. [\onlinecite{KharzeevPPNP2016}] for a review) which is generated in both the magnetic and electric fields. 
For the CSE to occur, one needs a finite chemical potential, while for CESE a chiral chemical potential is required. 
It is technically challenging to observe these effects in high-energy experimental setups.

In this paper we propose WSMs to be the platforms for the CSE and CESE observation. 
In some WSMs the chirality of the fermion can be formed with its spin, for example by spin-orbit coupling.
Because of that the chirality current generated through the CSE and CESE will be the spin current, and therefore can be measured by the inverse spin Hall effect of Kerr effect. 
In this paper we focus on the CESE because, as it is shown here, it has a very rich physical structure. 
We only discuss the chiral anomaly contribution to the CESE. 
In order to extract the chiral anomaly contriubtion under controlled approximation, we use a model of WSM in which the two valleys of opposite chirlities split in momentum.
Firstly, we show that an application of both the magnetic and electric fields is required for the CESE to be observed. 
There the role of the magnetic and electric fields is to stabilize a finite chiral chemical potential through a mechanism of the chiral anomaly. \cite{FukushimaKharzeevWarringaPRD2008, SpivakAndreevPRB2016, Yip, ZyuzinWSM}

Secondly, our calculations show that a general expression for the CESE has a number of components. 
Analogously to the terminology of the magnetoconductivity tensor, one can distinguish \textit{longitudinal}, and two transverse, \textit{in-plane transverse} and \textit{Hall}, components.
First two components are proportional to second power of the electric field and are linear in the magnetic field.
The in-plane transverse component depends on the angle $\theta$ between electric and magnetic fields as $\propto \sin(\theta)\cos(\theta)$ if measured in direction of the magnetic field. 
Such measurement geometry is justified as the main component of the chirality current, which is due to the CSE, is along the magnetic field.
Longitudinal component is proportional to $\cos^{2}(\theta)$. 
Hall component is quadratic in both the electric and magnetic fields, and depends on the angle in the same way as the in-plane transverse does. 
We note that, in contrast to the anomalous Hall effect in WSMs, we have not found any anomalous CESE due to the splitting of the Weyl points in momentum.

\section{Model of Weyl semimetal}
We study a model of Weyl semimetal, described by the following Hamiltonian,
\begin{align}\label{ham}
H_{s} =  s v \left[ \sigma_{x} k_{x} + \sigma_{y} k_{y} + \sigma_{z} k^{(s)}_{z}  \right] - \mu,
\end{align}
where $s=\pm$ denoting the two chiralities, $v$ is the velocity, $\mu \equiv vk_{\mathrm{F}}> 0$ is the chemical potential at zero temperature, and $k^{(s)}_{z} = k_{z} + sQ$, where $Q$ is the momentum separation of the Dirac nodes. 
Pauli matrices $\sigma_{i}$ correspond here to fermion $\mathrm{SU}(2)$ degree of freedom (e.g. spin, orbit, atom of the unit cell or their mixture).
The role of $Q$ in the following considerations is to only set an assumption about the impurity scattering times, and we omit it until mentioned. 
Spectrum is linear, $\epsilon_{{\bf k}\eta}  = \eta v k - \mu$, where $\eta = \pm$ is denoting conduction and valence bands.
We wish to study the responses of the system to the electric ${\bf E}$ and magnetic ${\bf B}$ fields.
We assume that all non-equilibrium responses are coming from the vicinity of the Fermi surface. 
In the following we restrict operations with the kinetic equation to the $\eta = +$ fermions, and omit $\eta$ there.
For equilibrium responses, on the other hand, $\eta = -$ is also important, and we restore the $\eta$ index in all relevant steps. 
To study non-equilibrium responses, we adopt a method of the kinetic equation for the distribution function of fermion state described by a wave-packet,
\begin{align}\label{kinetic_eq}
\frac{\partial n^{(s)}_{{\bf k}}}{\partial t}
+ {\dot {\bf k}}^{(s)} \frac{\partial n^{(s)}_{{\bf k}}}{\partial {\bf k}}
+ {\dot {\bf r}}^{(s)} \frac{\partial n^{(s)}_{{\bf k}}}{\partial {\bf r}}
= I_{\mathrm{coll}}\left[ n^{(s)}_{{\bf k}} \right],
\end{align}
supplimented with semi-classical equations of motion for the wave-packet \cite{XiaoChangNiuRMP2010},
\begin{align}
&
{\dot {\bf r}}^{(s)}_{\eta} = \frac{\partial \epsilon^{(s)}_{\eta{\bf k}} }{\partial {\bf k}}
+{\dot {\bf k}}^{(s)}_{\eta}\times {\bf \Omega}^{(s)}_{\eta{\bf k}},
\nonumber
\\
&
 {\dot {\bf k}}^{(s)}_{\eta} = e{\bf E} + \frac{e}{c}{\dot {\bf r}}_{\eta}^{(s)}\times{\bf B},
\end{align}
where, ${\bf \Omega}^{(s)}_{\eta{\bf k}} = -s\eta \frac{ {\bf k}}{2 k^3}$ is the Berry curvature, and note that the spectrum $\varepsilon^{(s)}_{\eta {\bf k}} = \epsilon^{(s)}_{\eta{\bf k}} - {\bf m}^{(s)}_{\eta{\bf k}}{\bf B}$ is updated by the orbital magnetization, ${\bf m}^{(s)}_{\eta{\bf k}} = -se\frac{v{\bf k}}{2ck^2}$, of the wave-packet. 
Both, the Berry curvature and the orbital magnetization, were shown to originate in the equations of motion after integrating out the valence band fermions. \cite{SonYamamotoPRD2013, GorbarPRL2017} 
See Ref. [\onlinecite{XiaoChangNiuRMP2010}] for the original derivation based on the wave-packet description. 
One solves the equations of motion,
\begin{align}\label{semiclassical}
& {\dot {\bf r}}^{(s)}_{\eta} = \frac{1}{\Delta^{(s)}_{\eta{\bf k}}}
\left[ {\bf v}_{\eta}^{(s)} + e {\bf E}\times{\bf \Omega}^{(s)}_{\eta{\bf k}} + \frac{e}{c}({\bf \Omega}^{(s)}_{\eta{\bf k}} {\bf v}^{(s)} ){\bf B} \right]
\nonumber
\\
&
 {\dot {\bf k}}_{\eta}^{(s)} = \frac{1}{\Delta^{(s)}_{\eta{\bf k}}}
\left[e{\bf E} + \frac{e}{c}{\bf v}^{(s)}_{\eta}\times {\bf B} + \frac{e^2}{c}\left({\bf E} {\bf B} \right){\bf \Omega}^{(s)}_{\eta{\bf k}} \right],
\end{align}
where $\Delta^{(s)}_{\eta{\bf k}} = 1 + \frac{e}{c}({\bf B} {\bf \Omega}^{(s)}_{\eta{\bf k}})$, and ${\bf v}^{(s)}_{\eta} = \frac{\partial \epsilon^{(s)}_{\eta{\bf k}}}{\partial \bf k} =  \eta v\frac{{\bf k}}{k} [ 1+ \frac{2e}{c} ({\bf B}{\bf \Omega}^{(s)}_{\eta {\bf k}} ) ] + s \frac{ev}{2c k^2}{\bf B}$.
Collision integral is only given by electron-impurity scattering,
\begin{align}
I_{\mathrm{coll}}\left[ n^{(s)}_{{\bf k}} \right] = 
- \int_{{\bf k}^{\prime}} \Delta_{{\bf k}^{\prime}}^{(s^\prime)} \omega^{(ss^\prime)}_{{\bf k}^\prime,{\bf k}}  
\left[ n^{(s)}_{{\bf k}} - n^{(s^\prime)}_{{\bf k}^{\prime}} \right]
\delta( \epsilon^{(s)}_{{\bf k}} - \epsilon^{(s^\prime)}_{{\bf k}^{\prime}}).
\end{align}
We follow the approximations and the formalism of treating the collision integral presented by the author in Ref. [\onlinecite{ZyuzinWSM}].
Namely, we consider a model of short-range impurities, introduce two scattering times, intra-chirality time $\tau_{\bf k}$ and inter-chirality scattering time $\tau_{\mathrm{V}}$, and assume the times to be isotropic, i.e. angle independent.\cite{comment} 
These approximations allow one to analytically extract chiral anomaly contribution to the distribution function. \cite{ZyuzinWSM} 
We note, in general, the scattering times are angle and magnetic field dependent, in which case the distribution function must be expanded in harmonics in order to approximate the kinetic equation.   
The collision integral is then
\begin{align}
I_{\mathrm{coll}}[ n^{(\pm)}_{{\bf k}} ]
= \frac{{\bar n}^{(\pm)} - n^{(\pm)}_{{\bf k}}}{\tau_{\bf k}}
+ \frac{{\bar n}^{(\mp)} - n^{(\pm)}_{{\bf k}}}{\tau_{\mathrm{V}}},
\end{align}
where we integrated over the angles and denoted ${\bar n}^{(s)} = \int_{\phi_{\bf k}\theta_{\bf k}} \Delta_{\bf k}^{(s)} n^{(s)}_{\bf k}$.
The collision integral is conveniently rewritten as
$
I_{\mathrm{coll}}[ n^{(s)}_{{\bf k}} ] = \frac{1}{\tau^{*}_{\bf k}} [ {\bar n}^{(s)} - n^{(s)}_{{\bf k}} ] + \Lambda^{(s)}
$, where an important quantity $\Lambda^{(\pm)} = \frac{1}{\tau_{\mathrm{V}}} [  {\bar n}^{(\mp)} - {\bar n}^{(\pm)} ]$ was introduced, and $\tau^{*}_{\bf k} =  ( \tau^{-1}_{\bf k} + \tau_{\mathrm{V}}^{-1} )^{-1}$ is the total scattering time. It is important to keep the ${\bf k}$ dependence of the intra-valley scattering time for consideration of the effects, as advertised in the Introduction to the paper, non-linear in electric field. 
The ${\bf k}$ dependence of the inter-valley scattering time can be ignored due to the splitting of the Weyl points in momentum space.  
Following the lines of Ref. [\onlinecite{ZyuzinWSM}] we multiply the kinetic equation by $\Delta_{\bf k}^{(s)}$, assume that there is a steady state, and integrate over the angles, and get
\begin{align}\label{chiralanomaly}
\int_{\phi_{\bf k}\theta_{\bf k}}\Delta_{\bf k}^{(s)}{\dot {\bf k}}^{(s)} \frac{\partial n_{{\bf k}}^{(s)} }{\partial {\bf k}} = \Lambda^{(s)},
\end{align}
where quantity on the left-hand side is
\begin{align}\label{mu5}
&\Lambda^{(s)}_{\mathrm{E}} \equiv - s  \frac{ e^2 v^2  }{6ck}\left( {\bf E}{\bf B}\right)
\left( \frac{2}{vk}\frac{\partial f}{\partial \epsilon_{\bf k}} - \frac{\partial^2 f}{\partial \epsilon_{\bf k}^2}\right),
\end{align}
where $\epsilon_{\bf k} = vk - \mu$ was introduced, and it was expanded in magnetic field where needed.
This is the quantity associated with the chiral anomaly, i.e. non-conservation of the chiral charge when the electric and magnetic fields are present.
We elaborate this point.
For example, if we would have assumed $\tau_{\mathrm{V}} = \infty$ then $ \Lambda^{(s)} = 0$, and there would be no steady state in the system. Then, after integrating the kinetic equation over ${\bf k}$ we obtain $\frac{\partial N^{(s)}}{\partial t} + {\bm \nabla} {\bf J}^{(s)} = \int  \Lambda^{(s)}_{\mathrm{E}} \frac{ k^2 dk}{2\pi^2} = s\frac{e^2}{4\pi^2 c}\left( {\bf E}{\bf B}\right)$ - non-conservation of the chiral charge, where $N^{(s)} = \int {\bar n}^{(s)}\frac{k^2 dk}{2\pi^2}$ and ${\bf J}^{(s)}$ is the local current. 
Obtained above equality (\ref{chiralanomaly}), $\Lambda^{(s)}_{\mathrm{E}} = \Lambda^{(s)}$, means that in order to have an assumed steady state, the collision integral must compensate the term due to the chiral anomaly.
This is the mechanism of the chiral chemical potential $\mu_{5}$ stabilization. \cite{FukushimaKharzeevWarringaPRD2008, SpivakAndreevPRB2016,Yip,ZyuzinWSM}
An estimate of the chiral chemical potential gives $\mu_{5} = \mu_{+} - \mu_{-} = -\tau_{\mathrm{V}}\frac{e^2 v}{3ck_{\mathrm{F}}^2}({\bf E}{\bf B})$.

\section{Chiral separation effect}
We first reproduce known results for the equilibrium chirality current when only the magnetic field is applied - the so-called chiral separation effect (CSE) \cite{SonZhitnitskyPRD2004, MetlitskiZhitnitskyPRD2005} and for a review see Ref. [\onlinecite{ KharzeevPPNP2016}].
It is known that CSE does not depend on the electron-impurities scattering and is topological in nature.
The chirality current is defined as
\begin{align}\label{chiralcurrent}
{\bf j}^{\mathrm{C}} = {\bf j}^{(+)} - {\bf j}^{(-)}.
\end{align}
As mentioned above, in equilibrium all states, i.e. $\eta = \pm$, contribute to the current.
Current associated with the chirality $s=\pm$ that contributes to the CSE is 
\begin{align}\label{current}
{\bf j}^{(s)} = \sum_{\eta}{\bf j}^{(s)}_{\eta}  =  \int_{{\bf k}}  \Delta_{\eta \bf k}^{(s)}{\dot {\bf r}}_{\eta}^{(s)}n^{(s)}_{\eta {\bf k}} ,
\end{align}
i.e. with the current corresponding to the anomalous Hall effect absent. 
In the wave-packet approach the magnetic field ${\bf B}$ enters equations in various places, and in order to get known numerical coefficient for the chirality current correct \cite{MetlitskiZhitnitskyPRD2005, KharzeevPPNP2016}, one needs to keep track of a number of terms in the current.
In equations of motions we set ${\bf E} = 0$, and to the first order in ${\bf B}$ we get an expression for the current 
\begin{align}
{\bf j}^{(s)}_{\eta} = \eta \frac{ev}{c}\int
 \frac{{\bf k}}{k} ({\bf B}{\bm \Omega}_{\eta \bf k}^{(s)}) 
 n_{\eta \bf k}^{(s)}
- \eta v\int\frac{{\bf k}}{k} ({\bf m}_{\eta \bf k}^{(s)}{\bf B}) \frac{\partial n_{\eta \bf k}^{(s)}}{\partial \epsilon_{\eta \bf k}}.
\end{align}
Integrating second term by parts, and then over the angles, we get
$
{\bf j}^{(s)}_{\eta} = -s\frac{ev}{4\pi^2 c}{\bf B}\int n_{\eta \bf k}^{(s)} dk.
$
Summing over all states, we get for the chirality current a known result \cite{MetlitskiZhitnitskyPRD2005, KharzeevPPNP2016} of
\begin{align}\label{cse}
{\bf j}^{\mathrm{C}}  = -\frac{e \mu}{2\pi^2 c}{\bf B} \equiv -\sigma_{\mathrm{B}}{\bf B},
\end{align}
where sign here is due to the definition of the Berry curvature. It is illuminating that we would have obtained the same result by ignoring the orbital magnetization altogether.

\section{Chiral electric separation effect due to the chiral anomaly}
We now study a situation when the electric field ${\bf E}$ is also applied to the system. 
The electric field will drive the fermion system out of equilibrium in the vicinity of the Fermi level. 
As discussed above, the intra-valley scattering processes will result in the average of the distribution function within the valley.
Compensation of the chiral anomaly by the inter-valley scattering processes will result in the chiral chemical potential proportional to $({\bf EB})$. 
As shown below, due to that there will be chiral anomaly contributions to the chirality current (\ref{chiralcurrent}).

We restrict the calculations of the responses to the conduction band $\eta = +$, omitting the $\eta$ index in the following. 
When approximating the kinetic equation we will be guided with a $\tau_{\mathrm{V}} \gg \tau^{*}_{\bf k}$ assumption, which is the case of a large separation $Q$ between the Weyl points.
We will be searching for contributions that are determined by the chiral anomaly mechanism, i.e. by $\Lambda_{\mathrm{E}}^{(s)} \propto \tau_{\mathrm{V}} ({\bf E}{\bf B})$ - ingredients for the chiral chemical potential in our model.
We expand the distribution function in powers of electric field, $n_{\bf k}^{(s)} \approx n_{{\bf k}, 0 }^{(s)} + n_{{\bf k}, 1 }^{(s)} + n_{{\bf k}, 2 }^{(s)} +..$, where introduced lower index corresponds to the power of the electric field. 
Using the expansion, we pick from the kinetic equation
$
 n^{(s)}_{{\bf k}}   
={\bar n^{(s)}} + \tau^{*}_{\bf k} 
[   \Lambda^{(s)}
- {\dot {\bf k}}^{(s)} \frac{\partial n_{{\bf k}}^{(s)}}{\partial {\bf k}} ]
$
only a
$n_{{\bf k}, 2 }^{(s)} \approx - \tau^{*}_{\bf k} e{\bf E}\frac{\partial n_{{\bf k}, 1 }^{(s)}}{\partial {\bf k}}$ term. 
It can be shown that all linear in ${\bf E}$ terms in the chirality current vanish.
We next substitute it to the expression for the $s=\pm$ current, 
$\delta{\bf j}^{(s)}_{\mathrm{E}} 
\approx v \int_{{\bf k}} \frac{{\bf k}}{k}  n_{{\bf k}, 2 }^{(s)}$
and pick the chiral anomaly contribution to the chirality current ${\bf j}^{\mathrm{C}}_{\mathrm{E}}$ as ${\bf E}\frac{\partial n_{{\bf k}, 1 }^{(+)}}{\partial {\bf k}} - {\bf E}\frac{\partial n_{{\bf k}, 1 }^{(-)} }{\partial {\bf k}} \approx - \tau_{\mathrm{V}}({\bf E} \frac{\bf k}{k})\frac{\partial \Lambda^{(+)}}{\partial k} $, with a $n_{{\bf k}, 1 }^{(s)} \approx \bar{n}_{{\bf k}, 1 }^{(s)}$ approximation.
Finally we arrive at 
\begin{align}
\delta {\bf j}^{\mathrm{C}}_{\mathrm{E}} &
= 
\tau_{\mathrm{V}}  \frac{e v}{6\pi^2} {\bf E}\int dk~ \tau^{*}_{\bf k}  k^2 \frac{\partial \Lambda^{(+)}}{\partial k}  
\nonumber
%\\
%&
%=
%-\tau_{\mathrm{V}}  \frac{e v}{6\pi^2} {\bf E}\int dk \left(\frac{\partial}{\partial k} \tau^{*}_{\bf k}  k^2  \right) \Lambda^{(+)}
%\nonumber
\\
&  \label{result1}
= -  \tau_{\mathrm{V}}  \tau_{\mathrm{F}} I_{1} 
\frac{e^3 v^2}{36 \pi^2 c \mu} 
({\bf E}{\bf B}) {\bf E} ,
\end{align}
where $I_{1} = \frac{k_{\mathrm{F}}}{\tau_{\mathrm{F}}} \left[ (\frac{\partial}{\partial k}  \tau^{*}_{\bf k}  k^2)\frac{1}{k^{2}}+(\frac{\partial^2}{\partial k^2}  \tau^{*}_{\bf k}  k^2)\frac{1}{k}\right]\vert_{k=k_{\mathrm{F}}}$ is a dimensionless quantity, where $\tau_{\mathrm{F}} = \tau_{{\bf k}}\vert_{k=k_{\mathrm{F}}}$ - intra-valley scattering time estimated at the Fermi momentum. For short-range impurities $\tau_{\bf k} = \tau_{\mathrm{F}}k_{\mathrm{F}}^2/ k^{2}$ and for $\tau_{\mathrm{V}} \gg \tau_{\mathrm{F}}$ we get $I_{1}\approx - 4\frac{\tau_{\mathrm{F}}}{\tau_{\mathrm{V}}}$, for Coulomb impurities $\tau_{\bf k} \equiv \tau_{\mathrm{F}} = \mathrm{const}({\bf k})$ and for $\tau_{\mathrm{V}} \gg \tau_{\mathrm{F}}$ we get $I_{1} \approx 4$.

Let us now search for the Hall type response of the chirality current.
For that we search for a correction to the distribution function due to the Lorentz force using the Zener-Jones method. \cite{ZenerJonesPRSLA1934, PalMaslovPRB2010}
Same expansion in electric field as above gives a correction of interest,
\begin{align}
n_{{\bf k}, 2 }^{(s)} \approx \frac{e^2(\tau^{*}_{\bf k})^2}{c} ({\bf v}\times{\bf B})\frac{\partial}{\partial {\bf k}}\left[ {\bf E}
\frac{\partial n_{{\bf k}, 1 }^{(s)}}{\partial {\bf k}}\right],
\end{align}
with which we calculate the chirality current in the same way as was done above. 
In calculating it, an identity $({\bf v}\times{\bf B})\frac{\partial}{\partial {\bf k}} g(\vert {\bf k} \vert) = 0$, where $g(x)$ is an arbitrary function, is of use. 
We get for the chirality current, 
\begin{align}\label{result2}
\delta {\bf j}^{\mathrm{C}}_{\mathrm{H}} 
=  \tau_{\mathrm{V}}\tau_{\mathrm{F}}^2 I_{2}\frac{e^4 v^4}{36\pi^2 c^2 \mu^2}  ({\bf E}{\bf B}) 
[{\bf B}\times{\bf E}],
\end{align}
where $I_{2} = \frac{k_{\mathrm{F}}^2}{\tau_{\mathrm{F}}^2} 
\left[ (\frac{\partial}{\partial k}  (\tau^{*}_{\bf k})^2  k)\frac{1}{k^{2}}+(\frac{\partial^2}{\partial k^2}  (\tau^{*}_{\bf k})^2  k)\frac{1}{k}\right]\vert_{k=k_{\mathrm{F}}}$ is dimensionless quantity.  For short-range impurities $\tau_{\bf k} = \tau_{\mathrm{F}}k_{\mathrm{F}}^2/ k^{2}$ and for $\tau_{\mathrm{V}} \gg \tau_{\mathrm{F}}$ we get $I_{2}\approx 9$, for Coulomb impurities $\tau_{\bf k} \equiv \tau_{\mathrm{F}} = \mathrm{const}({\bf k})$ and for $\tau_{\mathrm{V}} \gg \tau_{\mathrm{F}}$ we get $I_{2} \approx 1$.

\begin{figure}[t] \centerline{\includegraphics[clip, width=0.8  \columnwidth]{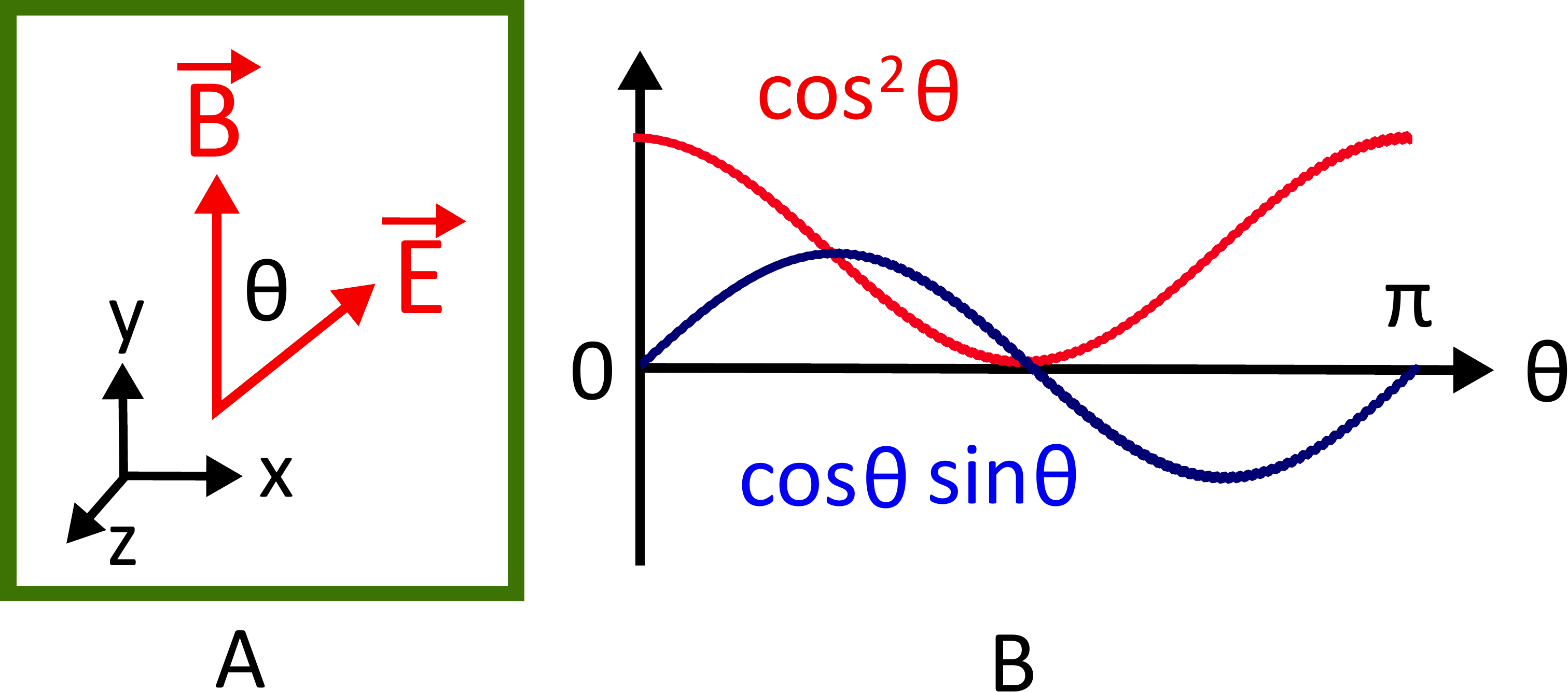}}

\protect\caption{(Color online) Schematics of the CESE. 
(A). Configuration of the electric and magnetic fields: magnetic field {\bf B} is chosen along $y-$direction, while electric field ${\bf E}$ is in the $xy$ plane making an angle $\theta$ with the magnetic field. The chirality current given by $j^{\mathrm{C}}_{\alpha} = \sigma_{\alpha\beta}B_{\beta}$ can flow in all three directions, with the components called: $\sigma_{yy}$ - CSE with a correction due to the \textit{longitudinal} CESE,  $\sigma_{xy}$ - \textit{in-plane transverse} CESE, and $\sigma_{zy}$ - \textit{Hall} CESE. 
(B). Characteristic $\cos^{2}\theta$ angle dependence of the longitudinal CESE is shown in red, while dependence of the in-plane transverse and Hall CESE both obeying the $\cos\theta \sin\theta$ dependence is shown in blue.}

\label{fig:plot}  

\end{figure}

Derived contributions to the chirality current (\ref{result1}) and (\ref{result2}) is the chiral electric separation effect (CESE). 
Total chirality current with CESE parts due to the chiral anomaly is ${\bf j}^{\mathrm{C}} = - \sigma_{\mathrm{B}}{\bf B} -  \chi I_{1} ({\bf E}{\bf B})  {\bf E} 
+ (\omega_{\mathrm{c}}\tau_{\mathrm{F}}) \chi I_{2}  ({\bf E}{\bf B})  \frac{[{\bf B}\times{\bf E}]}{B}$, and for large $\mu$ the first term will dominate over the other two. 
In analogy with regular magnetoconductivity terminology, we wish to introduce description of the components of the CESE based on the measurement in the direction of the magnetic field rather than electric.
It is natural to break the chirality current in to components, such that $j^{\mathrm{C}}_{\alpha} = \sigma_{\alpha\beta}B_{\beta}$, where
\begin{align}
&
\sigma_{yy} = - \sigma_{\mathrm{B}} - \chi I_{1}  E^2 \cos^{2}(\theta),
\nonumber
\\
&
\sigma_{xy} =  - \chi I_{1}  E^2 \cos(\theta)\sin(\theta),
\\
&
\sigma_{zy} =  (\omega_{\mathrm{c}}\tau_{\mathrm{F}} )\chi I_{2}  E^2  \cos(\theta)\sin(\theta),
\nonumber
\end{align}
where $\omega_{\mathrm{c}}  = \frac{eB v}{ck_{\mathrm{F}}}$ is the cyclotron frequency, and $\chi = \tau_{\mathrm{V}}\tau_{\mathrm{F}}\frac{e^3 v^2}{36\pi^2 c \mu}$ is introduced for convenience.
Therefore, the components can be called as: $\propto E^2$ part of the $\sigma_{yy}$  - \textit{longitudinal}, $\sigma_{xy}$- \textit{in-plane transverse}, and $\sigma_{zy}$ is the \textit{Hall} component of CESE.  
Interestingly, the Hall component depends on the relative angle between electric and magnetic fields $\theta$ in the same way as the in-plane transverse does, namely $\propto \cos(\theta)\sin(\theta)$. See Fig. (\ref{fig:plot}) for schematics.

We have checked that contrary to the charge transport, where one gets the anomalous Hall effect if the two Weyl points are split in momentum, there is no anomalous response for the chiral current due to the splitting.

Formally, the obtained above results, Eqs. (\ref{result1}) and (\ref{result2}), are valid in case $\frac{\omega_{\mathrm{c}}}{\mu} < 1$ - condition used in deriving the semiclassical equations of motion updated by the Berry curvature, and $\omega_{\mathrm{c}}\tau_{\mathrm{F}} < 1$, where $\tau_{\mathrm{F}}$ is the intra-valley relaxation time estimated at the Fermi energy - condition used in the Zener-Jones method which we have applied for the extraction of the chiral Hall effect. 
In deriving the results, we have assumed $\tau_{\mathrm{V}} \gg \tau_{\mathrm{F}}$ - a valid approximation for the Weyl semimetal, where the two valleys are split in momentum.
Another approximation is $\vert\mu_{5}\vert = \tau_{\mathrm{V}}\frac{e^2 v}{3ck_{\mathrm{F}}^2}\vert ({\bf E}{\bf B}) \vert < \mu$, which states that the conduction bands for both valleys are not empty and we can use the kinetic equation with the valence bands ignored.
Other than that the approximation of the kinetic equation is in line with the linear response in electric field. 
Namely, in Eq. (\ref{result1}), the $({\bf E}{\bf B})$ part came from the $n_{+}- n_{-} \propto \tau_{\mathrm{V}}({\bf E}{\bf B})$ (chiral anomaly), which is an unperturbative expression, and another ${\bf E}$ came from the equation of motion, i.e. $\dot{\bf{k}}$, within the linear response formalism. 
Same applies for the obtained Eq. (\ref{result2}).
We can, therefore, claim that in a two valley fermion system, kinetic equation can be used to extract the non-linear corrections to the chirality current.

We now comment on possible realization of the proposed effect. 
The chirality current can be observed through a possible connection of the chirality with the spin of fermions.
If in the Hamiltonian, given by Eq. (\ref{ham}), $i$th Pauli matrix corresponds to a real spin of fermions, the current carried by a given chirality in the $i$th direction will be spin polarized. 
Such that fermions with opposite chiralities will carry opposite spin, and the CESE will formally correspond to the spin current generated in the $i$th direction. The spin current can be measured using the inverse spin Hall effect or by the Kerr rotation of light on the edges of the system where the spin will be accumulated.

\section{Conclusions}
To conclude, we have considered the chiral electric separation effect (CESE) in Weyl semimetals.
Our main results are represented by equations (\ref{result1}), (\ref{result2}). 
Eq. (\ref{result1}) describes longitudinal and in-plane transverse components of CESE, while Eq. (\ref{result2}) is the Hall component. 
All three components are due to the chiral anomaly.
Novelty of the results is in their unusual dependence on the electric and magnetic fields, and the angle between them.

In WSMs the CSE and CESE can be associated with the spin currents, and can be experimentally measured.
Therefore, according to the obtained results, one can manipulate with the spin currents using both the electric and magnetic fields - an essence of the spintronics research field. \cite{DyakonovBOOK2008} 
Furthermore, if measured together with the positive LMC and the in-plane TMC, the effects might serve as another indication of the chiral anomaly mechanism behind the positive LMC and in-plane TMC in WSMs. \cite{NielsenNinomiyaPLB1983, SonSpivakPRB2013, SpivakAndreevPRB2016, Yip, ZyuzinWSM}
This is because the positive LMC, in-plane TMC and CESE originate from the same physics of the chiral chemical potential stabilization.

We note that Eqs. (\ref{result1}), (\ref{result2}) agree with consideration of the CESE for chiral plasma given in Ref. [\onlinecite{GorbarCESE2016}]. 
There, given chiral chemical potential is known, all possible contributions to CESE are listed (see Eq. (72) of Ref. [\onlinecite{GorbarCESE2016})].
In the present paper the chiral chemical potential is found to be generated when the electric and magnetic fields are applied, and its expression is estimated (discussion after Eq. (\ref{mu5})). 
If we put the estimated $\mu_{5} = -\tau_{\mathrm{V}}\frac{e^2 v}{3ck_{\mathrm{F}}^2}({\bf E}{\bf B})$ in to Eq. (72) of the Ref. [\onlinecite{GorbarCESE2016}] and put $v = c$, we get a reasonable match between the results.

We note that similar effect to CSE was recently proposed for photons in Refs. [\onlinecite{YamamotoPRD2017, ZyuzinPRA2017, ZyuzinPRA2017comment}]. 
There the role of the magnetic field is the rotation of the material - vorticity. 
We hope that effects similar to the ones proposed in the present paper will also be found for photons.

VAZ is thankful to A.A. Zyuzin for pointing out the importance of ${\bf k}-$ dependence of the intra-valley scattering time for the CESE description. 
Extension of the present paper to the AC electric field is given in Ref. [\onlinecite{ZMZ}].
The author thanks Landau Institute for Theoretical Physics and Pirinem School of Theoretical Physics for hospitality.

\end{document}